\documentclass[a4paper,10pt]{extarticle}

\usepackage{preprint}
    
\usepackage{amssymb}
\usepackage{epsfig}
\usepackage{graphicx}
\usepackage{times}
\usepackage{float}
\usepackage{xspace}
\usepackage{xfrac}
\usepackage{enumerate}
\usepackage{url}
\usepackage{subcaption}
\usepackage{pgfplots}
\pgfplotsset{compat=1.15}

\usepackage{algorithm}
\usepackage{algpseudocode}

\usepackage{listings}

\usepackage{makecell}

\usepackage{pgfplots}
\usepgfplotslibrary{external}
\usepgfplotslibrary{groupplots}

\usepackage{geometry}

\usepackage[acronym,nohypertypes={acronym}]{glossaries}
\newacronym{hpc}{HPC}{high performance computing}
\newacronym{mpi}{MPI}{message passing interface}
\newacronym{cfd}{CFD}{computational fluid dynamics} \newcommand{\cfd}{\gls{cfd}\xspace}
\newacronym[shortplural=LES]{les}{LES}{large eddy simulation}
\newacronym{rans}{RANS}{Reynolds-averaged Navier-Stokes}
\newacronym{rsm}{RSM}{Reynolds stress model}
\newacronym{abl}{ABL}{atmospheric boundary layer}
\newacronym{lbm}{LBM}{lattice-Boltzmann method} \newcommand{\lbm}{\gls{lbm}\xspace}
\newacronym{clbm}{CLBM}{cumulant lattice-Boltzmann method}
\newacronym[plural=PDFs]{pdf}{PDF}{particle distribution function}
\newacronym{alm}{ALM}{actuator line model} \newcommand{\alm}{\gls{alm}\xspace}
\newacronym{adm}{ADM}{actuator disk model}
\newacronym{bem}{BEM}{blade element momentum}
\newacronym[plural=TSRs]{tsr}{TSR}{tip-speed-ratio}
\newacronym{api}{API}{application programming interface}
\newacronym{aep}{AEP}{annual energy production}
\newacronym{hawt}{HAWT}{horizontal axis wind turbine}
\newacronym{vawt}{VAWT}{vertical axis wind turbine}

\newacronym{gpu}{GPU}{graphics processing unit}
\newacronym{mlups}{MLUPS}{\textsl{Mega Lattice-Site Updates per Second}}
\newacronym{flop}{FLOP}{floating-point operation}
\newacronym{flops}{FLOPS}{floating-point operations per second}
    
\newcommand{\walberlawind}{\mbox{\textsc{waLBerla-wind}}\xspace}
\newcommand{\walberla}{\mbox{\textsc{waLBerla}}\xspace}
\newcommand{\lbmpy}{\textsc{lbmpy}\xspace}
\newcommand{\pystencils}{\textsc{pystencils}\xspace}
\newcommand{\castor}{\textsc{Castor}\xspace}

\newcommand{\nvcc}{\texttt{nvcc}\xspace}

\usepackage[capitalize]{cleveref}
\crefformat{figure}{\mbox{figure~#2#1#3}}
\Crefformat{figure}{\mbox{Figure~#2#1#3}}
\crefformat{section}{\mbox{section~#2#1#3}}
\Crefformat{section}{\mbox{Section~#2#1#3}}

\usepackage{fancyhdr}
\fancypagestyle{frontmatter}
{
   \fancyhead{}
   \fancyfoot[C]{\textcolor{Accent}{\small{Preprint}}}
}

\fancypagestyle{mainmatter}
{
    \fancyhead[R]{\footnotesize{\thepage}}
    \fancyhead[L]{}
    \fancyfoot[C]{\textcolor{Accent}{\small{Preprint}}}
    
}
	    
\begin{document}

\title{\walberlawind: a lattice-Boltzmann-based high-performance flow solver for wind energy applications}

\author{%
\textbf{Helen Schottenhamml\textcolor{Accent}{\textsuperscript{1,2,*}}, %
Ani Anciaux-Sedrakian\textcolor{Accent}{\textsuperscript{1}}, %
Fr\'{e}d\'{e}ric Blondel\textcolor{Accent}{\textsuperscript{1}}, %
Harald K\"{o}stler\textcolor{Accent}{\textsuperscript{2}}, %
Ulrich R\"{u}de\textcolor{Accent}{\textsuperscript{2,3}}}\\
\begin{small}
\textcolor{Accent}{\textsuperscript{1}}IFP Energies Nouvelles, Rueil-Malmaison, France \\ 
\textcolor{Accent}{\textsuperscript{2}}Friedrich-Alexander University, Erlangen, Germany \\
\textcolor{Accent}{\textsuperscript{3}}CERFACS, Toulouse, France \\ 
\textcolor{Accent}{\textsuperscript{*}}Correspondence: \textcolor{Accent}{\url{helen.schottenhamml@fau.de}} \\ \end{small}
}

\pagestyle{mainmatter}

\maketitle
\thispagestyle{frontmatter}

\begin{doublespacing}

\noindent
\begin{abstract}
This article presents the development of a new wind turbine simulation
software to study wake flow physics.
To this end, the design and development of \walberlawind{}, a new simulator based on the lattice-Boltzmann method that is known for its excellent performance and scaling properties, will be presented. 
Here it will be used for large eddy simulations (LES) coupled with actuator wind turbine models.
Due to its modular software design, \walberlawind{} is flexible and extensible with regard to turbine configurations. 
Additionally it is performance portable across different hardware architectures, another critical design goal.
The new solver is validated by presenting force distributions and velocity profiles and comparing them with experimental data and a vortex solver.
Furthermore, \walberlawind's performance is compared to a theoretical peak performance, and analysed with weak and strong scaling benchmarks on CPU and GPU systems.
This analysis demonstrates the suitability for large-scale applications and future cost-effective full wind farm simulations.

\end{abstract}


\section{Introduction} \label{sec:introduction}
\subsection{Overview}
\noindent
Climate change poses many challenges in today's politics, economics, and technology. 
The use of wind energy is a crucial step 
to decarbonize our future energy supply
\cite{irena2019}.
%
%
In this process, optimizing the design of single wind turbines and the geographical placement of a whole wind farm are crucial to achieve a maximal energy harvest. 
This process, called siting, eventually determines the net revenues of a farm, as well as environmental impacts and noise nuisance.
Once engineers have identified a suitable area for the wind farm, the exact placement of every single turbine on the site, the \mbox{micrositing}, remains a challenge as the turbines do not only interact with their environment but also with each other. 
Numerical simulations have become essential tools to maximize the \gls{aep} while minimizing unwanted effects and the overall cost per unit of energy.

Towards these goals, the current article will present the design and development of \walberlawind{}, a new simulation framework for wind turbines and wind farms based on the \lbm{}.
The approach is based on the \alm{}, where the blades of a turbine are respresented by surrogate models, as a compromise between physical fidelity and computational efficiency.
The \lbm{} is chosen because of its suitability for parallel execution on supercomputers, and its additional suitability for modern accelerator hardware, such as GPUs.
To our best knowledge, this article is the first to report on LBM wind energy simulations using parallel clusters with multiple GPUs. 

Besides a thorough assessment of the physical validity of the method, we will report on the software structure of \walberlawind{}.
Our emphasis is on techniques to achieve a flexible and user-friendly software
in a systematic development process that is suited for modern complex and heterogeneous supercomputer architectures. 
To achieve sustainability, we employ methods of advanced software engineering in the form of a clear modular software structure. 
In particular we employ an automatic generation of hardware-specific code from abstract specifications, so that both CPUs of GPUs can be used efficiently.
The success of these techniques will then be demonstrated by a careful scalability and performance analysis. 

The interdisciplinary combination of all topics, as reported in this paper, range from the development of the model and its validation to the realization in algorithms and software on advanced supercomputers.
This is an exemplary exercise in computational science and engineering (CSE) where all these aspects come together.  

\subsection{Wind energy}
Within the design process, the layout optimization of wind farms requires thousands of \gls{aep} estimations obtained by analytical wind farm flow models. 
While these models allow predicting the performance of a wind farm based on environmental conditions in a few seconds, they rely on sharp modeling assumptions such as neglecting viscosity and pressure terms in the continuity equations. 
Furthermore, such models contain several calibration constants that have to be determined. 
Additionally, new models are frequently developed to account for phenomena of interest, such as secondary steering or stability effects. 
To this end, they demand extensive databases for detailed flow analysis, calibration, and validation.\\
Different approaches exist to obtain such reference data sets. 
On the one hand, engineers may rely on experimental data from on-site or wind tunnel measurements. 
However, there are several issues with obtaining and using experimental data. 
Even though they provide great insight into the flow's details in scaled wind farms, wind tunnel measurements exhibit reduced flexibility in the setup and are often incapable of dealing with atmospheric stability. 
Usually, they only support neutral atmospheric conditions but do not account for stable and convective cases. On-site measurements, however, are very costly and subject to the randomness of the atmosphere, resulting in an inability to control the operating conditions. To this end, the measurements taken in real-world wind farms may be challenging to interpret.
%
\subsection{Related work in wind energy simulations}
The second big approach to setting up calibration and validation databases is using 
numerical solvers based on 
complex modeling approaches, including \gls{rans} simulations and \glspl{les}.
Despite constituting a reliable and accurate tool in many application fields, \gls{rans} underperforms in the context of wind farm flows \cite{steiner2022}. 
This deficiency is most probably due to the large velocity gradients observed in the wakes \cite{vanderlaan2013} and the anisotropy of \gls{abl} flows that traditional turbulence closures cannot represent correctly. 
Yet, it is worth mentioning that recent studies that use specialized \gls{rsm} approaches yield satisfying results of \gls{rans} simulations in wind turbine aerodynamics, e.g., \cite{baungaard2022}. 
On the other hand, \gls{les} simulations resolve the largest turbulent scales than \gls{rans} and provide the necessary degree of accuracy to model the observed phenomena in wake flows.
Consequently, high-fidelity solvers based on \gls{les} yield a viable alternative for constructing the databases. 
Solvers commonly used in the wind energy community, such as SOWFA \footnote{\url{https://www.nrel.gov/wind/nwtc/sowfa.html}} or \textsc{EllipSys3D} \cite{michelsen1992, michelsen1994, sorensen2003}, rely on the discretization of the Navier-Stokes equations and have been thoroughly tested and validated \cite{martinez-tossas2015,sorensen2002,troldborg10,sarlak2015}.\\
Whereas the traditional \gls{les} methods are convincing due to their superior accuracy in contrast to \gls{rans}, they require more computational resources and longer simulation runtimes, mainly due to their unsteady nature.
To reduce computational cost, many solvers refrain from fully resolving the geometry of the wind turbines and employ surrogates instead. 
A prominent example is actuator-based models such as the \gls{adm} or the \gls{alm}, first introduced by Sørensen et al. \cite{sorensen2002}. 
Here, wind turbines are represented only by the forces acting on the fluid calculated based on tabulated drag and lift data. 
This method has spread widely in wind farm aerodynamics and undergoes further improvement and extension. 
More recent advances, e.g., Churchfield et al. \cite{churchfield2017} discuss the impact and importance of a suitable regularisation kernel for force spreading. 
Additionally, so-called tip-loss models are introduced that add a correction factor to compensate for, among others, under-resolved tip vortices in the simulation \cite{mikkelsen2001,shen2005}.
Whereas the \gls{alm} drastically decreases the required computational resources while maintaining sufficient accuracy, the simulations remain expensive and show limited suitability for large-scale wind farms.\\
%
One alternative approach to the classical Navier-Stokes solvers for \gls{les} is the \lbm, a mesoscopic method with origin in statistical physics.
\lbm has gained increasing interest in the last years due to, among others, its impeccable performance and parallelization capabilities.
Despite being applied to a large variety of applications early on, it was more recently that the \lbm came into use in wind turbine aerodynamics. 
As for the classical Navier-Stokes solver, research first focused on geometrically resolved methods of wind turbine rotors and farms. 
Several groups, e.g., Pérot et al. \cite{perot2012} and Deiterding et al. \cite{deiterding2016}, showed the potential of \lbm for load and noise prediction, as well as wake propagation in the wind energy context.\\
However, it was only in 2018 that, to our best knowledge, \lbm was first combined with an \acrfull{alm} in \cite{rullaud2018}. 
Asmuth et al. then performed several studies on the accuracy \cite{asmuth2020general}, numerical sensitivity \cite{asmuth2019}, performance evaluation \cite{asmuth2019}, and compressibility effects \cite{asmuth2020compressibility} to assess the full potential of \lbm for wind turbine simulations. 
Notably, Asmuth et al. performed all these investigations on a single consumer \gls{gpu} card.\\
\subsection{Supercomputers with GPUs}
Modern parallel computing resources are built from complex hardware components and often use hybrid architectures. 
On the CPU side, they consist of thousands of multi-core processors, each consisting of several cores with a specific architecture. 
For parallelization, those systems use shared-memory and distributed-memory parallelism, as well as vectorization. 
The hybrid systems add further accelerating devices as general-purpose GPU cards, which, in turn, come with their individual hardware architecture.
To fully exploit a specific computational resource and reach outstanding performance, the software must address all levels of parallelism, as described above. 
%
In \cite{schottenhamml2022}, we recently introduced an \gls{alm} implementation based on the open-source \lbm solver \walberla \cite{schottenhamml2022}. 
This software is not restricted to a using single GPU but enables multi-node CPU and multi-GPU simulations. 
\cite{schottenhamml2022} outlined a first physical validation and preliminary performance investigations. 
The present article will present the software design in detail, including physical validation and performance analysis.
A particular emphasis is to deliver a flexible code base that is physically extensible and sustainable while achieving performance portability also to future supercomputer configurations.

\subsection{Content of the article}
To this end, the paper is structured as follows:
After a brief theoretical overview of the employed methods in Section \ref{sec:numerical_methods}, we will detail the software itself in Section \ref{sec:implementation}. 
Starting with the frameworks used as its fundament, i.e., the \lbm framework \walberla and the code generation frameworks \pystencils and \lbmpy, we will then detail our \alm implementation with a focus on its software design and parallelization strategy. 
In Section \ref{sec:results}, we will first revisit the physical validation and extend it by the \textsc{NewMexico} test case, and second, have a closer look at the performance and scalability on large-scale setups.


\section{Numerical Methods} \label{sec:numerical_methods}

%
\subsection{The lattice-Boltzmann Method} \label{sec:lbm}
\noindent
The \acrfull{lbm} \cite{chen1998} provides an efficient and modern alternative to classical solvers in computational fluid dynamics that rely on the discretization of the Navier-Stokes equations. 
Having its origin in lattice gas methods, the mesoscopic \gls{lbm} models the flow through the evolution of the \glspl{pdf} on a regular and uniform grid. 
In contrast to other methods, the \gls{lbm} also discretizes the velocity space. Consequently, information about the \glspl{pdf} can only propagate in a pre-defined direction $\mathbf{c}_i$. 
A $DdQq$ stencil describes a velocity set with $q$ distinct velocities $\mathbf{c}_i$ for a $d$-dimensional domain. 
In this work for wind turbine applications, we restrict ourselves to the $D3Q27$ stencil. 
It is accurate enough to capture the complex physics of the flow but only includes direct neighbors and, therefore, does not compromise the computational performance too much. 
For each velocity in a set, a cell stores one \gls{pdf} $f_i$ which represents the probability density of fluid particles moving from position $\mathbf{x}$ to $\mathbf{x}+\mathbf{c}_i\Delta t$ in a time step $\Delta t$. 
The link to the macroscopic variables is given by the moments of the \gls{pdf}. 
As such, the mass density and the momentum density for a fluid, subject to a force $\mathbf{F}$, are given as $\rho = \sum_i f_i$ and $\rho \mathbf{u} = \sum_i \mathbf{c}_i f_i + \frac{F \Delta t}{2 \rho}$, respectively. Finally, to predict the fluid flow, one considers the fully discrete lattice-Boltzmann equation 
\begin{equation}
f_i (\mathbf{x} + \mathbf{c}_i\Delta t, t + \Delta t) = f_i(\mathbf{x},t) + \Omega_i(\mathbf{x},t) + S_i(\mathbf{x},t).
\label{eq:lbm}
\end{equation}
Here, the source term $S_i$ contains the force contributions. 
Its precise form depends on the chosen force model. 
Here, we 
use the Guo force model \cite{guo2002}.
Typical implementations split up this update scheme into two parts. 
First, the collision step is dictated by the collision operator $\Omega_i$ and evaluates the right-hand side of \Cref{eq:lbm}. 
This operation is usually cell-local and hence well-suited for parallelization. 
The assignment of the \glspl{pdf} to the neighboring cells, then, is performed in the streaming step.\\
The choice of the collision operator, $\Omega_i$, is crucial for the stability and accuracy of the particular method. 
Often, moment-based collision operators are preferred due to their simplicity in implementation. 
However, especially for high-Reynolds number flows, these operators tend to have numerical instabilities. 
Several more sophisticated lattice-Boltzmann methods have been introduced in the last years to overcome this issue. 
In this work, we use the \gls{clbm} by Geier et al.~\cite{geier2015}. 
Instead of relaxing single populations or their moments, the \gls{clbm} operates on the populations' statistical observables, the cumulants.
Though it exhibits superior stability and accuracy, the method is considered to be algorithmically complex so that its implementation may therefore be difficult to optimize for better computational performance.


\subsection{Wind Turbine Modelling} \label{sec:turbine_model}
\noindent
The \gls{alm}, developed by Sørensen et al. \cite{sorensen2002}, is a useful tool to reduce the computational complexity and, therefore, increase the computational performance in wind turbine simulations. Instead of geometrically resolving the turbines, a surrogate only displays the forces acting on the fluid. In the following, we detail the implementation based on \cite{mikkelsen2001} and \cite{churchfield2015}.\\
The spatial discretization of wind turbines provides the first crucial step toward their coupling to the fluid field.
Roughly following the work of Joulin \cite{joulin2019}, we split a wind turbine into several components with different rotational and translational degrees of freedom. 
Every component consists of several discrete nodes that calculate the positions, velocities, and orientations at every time step. 
As their relative offset and orientation determine the elements' connection, geometrical dependencies are taken into account automatically. 
With this approach, arbitrary wind turbine configurations become feasible. Horizontal and vertical axis wind turbines, but also helicopter rotors and floating wind turbines, can be realized easily.
Given the spatial discretization, it remains to elaborate on how the acting forces are calculated and coupled to the fluid flow.
Actuator-type models are based on the blade-element theory and follow three steps. 
First, the density $\rho$ and velocity $\mathbf{u}$ are interpolated from the flow field at every discrete node. 
Even though different interpolation methods are possible, we use trilinear interpolation for both density and velocity.
After calculating the blade element forces $\mathbf{f}$, they are eventually projected back onto the flow field by taking the convolution with a regularization kernel $\eta_\epsilon$, i.e., $\mathbf{f}_\epsilon = \mathbf{f} \otimes \eta_\epsilon$. Often, the Gaussian regularisation kernel is chosen. 
However, in this work, we use the discrete Dirac delta kernel by Roma et al. \cite{roma1999}, avoiding the costly evaluation of a convolution product. 
Another advantage of this discrete kernel is its inherent conservation properties, i.e., in contrast to other discrete kernels, its sum of weights is always $1$, ensuring the conservation of force during the projection.\\
The calculation of the actual forces depends on the specific actuator-type model. 
In the case of the actuator line method, the local force acting on one blade element reads
\begin{equation}
F_L = \frac{1}{2} \rho u^2_{rel} w l (C_L \mathbf{e}_L + C_D \mathbf{e}_D),
\end{equation}
where $\mathbf{e}_L$ and $\mathbf{e}_D$ are the directions in which lift and drag forces act, respectively. 
In particular, $e_L$ points normal to the relative wind speed, whereas $e_D$ points in the direction of the relative wind speed. 
$\rho$ and $u_{rel}$ denote the interpolated density and the relative velocity at an actuator element of length $l$. 
For the modeling of blades, the chord length of the reference airfoil section defines the element width $w$. 
The lift and drag coefficients $C_L$ and $C_D$ depend on the local angle of attack $\alpha$ based on tabulated data.


\section{Implementation} \label{sec:implementation}


\subsection{waLBerla} \label{sec:walberla}
\noindent
The proposed wind turbine implementation is realized using the \walberla framework \cite{bauer2020}.
\walberla is an open-source multi-physics framework\footnote{\url{https://walberla.net}} with a focus on \gls{cfd} simulations with the lattice-Boltzmann method. Besides supporting the creation of reusable and maintainable applications through its modularity, \walberla targets exceptional performance and scalability across different hardware and software platforms.
To this end, \walberla employs domain-specific code generation techniques and a block-structured domain partitioning to allow for large-scale simulations on supercomputers. 
The block-structuredness, i.e., grouping the lattice cells into blocks of equal layouts, grants highly efficient kernels and communication schemes. 
The communication between neighboring blocks on different processes is realized via the \gls{mpi} and \textsl{ghost layers}, also called \textsl{halo layers}.
Combined with the partitioning of the domain, \walberla also provides routines for static and dynamic mesh refinement, serving the demands in many engineering applications. 
Following the block-structured approach, \walberla refines on the block level to preserve the lattice cell layout required for performance and scalability. 
For the optimal distribution of blocks among the processes, several load balancing schemes, e.g., space-filling curves, are available \cite{schornbaum2016,schornbaum2018}.


\subsection{lbmpy} \label{sec:lbmpy}
\begin{figure}[htbp]
\centering
\includegraphics[width=0.4\textwidth]{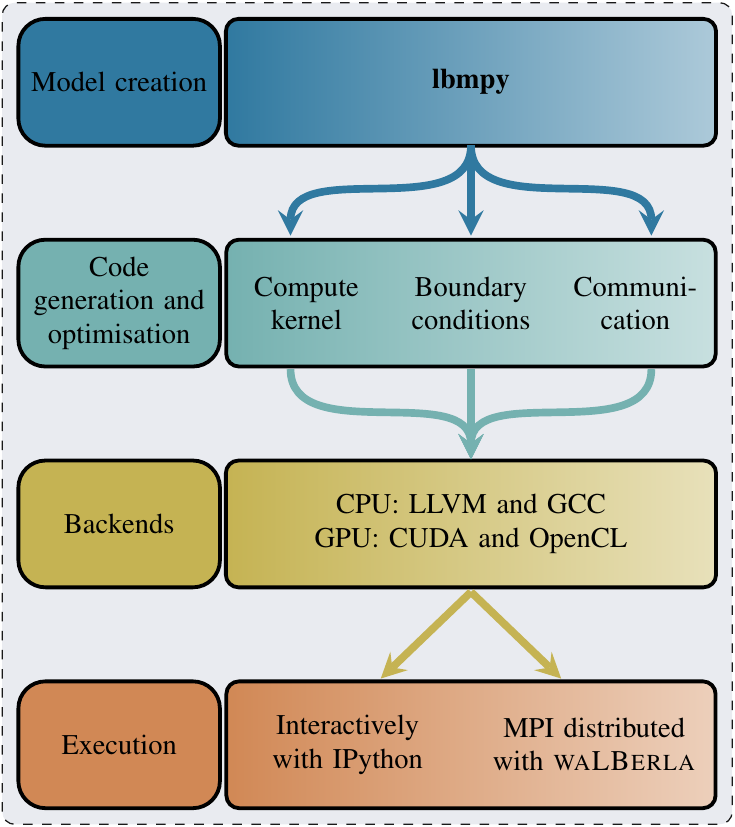}
\caption{Workflow for \walberla simulations using the code generation framework \lbmpy from Holzer et al. \cite{holzer2020}.}
\label{fig:lbmpy_workflow}
\end{figure}

\noindent
To further exploit the large-scale capabilities of \walberla, we utilize the open-source\footnote{\url{https://i10git.cs.fau.de/pycodegen/lbmpy}} code generation framework \lbmpy\cite{bauer2021} for the generation of \gls{lbm} collision and streaming schemes, communication, and boundary handling. By doing so, we can address conflicting software requirements for \gls{hpc} applications. 
On the one hand, we obtain maintainable, readable, and flexible code on a high-level basis.
On the other hand, we can still realize platform-specific performance optimizations.\\
For \gls{lbm} schemes, there are many possible combinations alone when considering the varieties of the method itself: the collision model, e.g., moment-based or cumulant-based, streaming patterns, force schemes, etc. 
But also with respect to the implementation, there are different options regarding storage patterns and hardware-dependent optimizations, such as loop-splitting and blocking. 
Instead of hand-optimizing a limited number of combinations, \lbmpy approaches this challenge with the symbolic representation of models and the automated generation of the corresponding kernels.\\
\lbmpy consists of several layers of abstractions based on Python's computer algebra package \textsc{sympy}. 
Expressing the models via \textsc{sympy} expressions allows the application developer to formulate the methods in a concise mathematical manner. 
In the following, we will only briefly address the workflow for combining \walberla and \lbmpy following \Cref{fig:lbmpy_workflow}. 
For a more detailed treatment, we refer to, Bauer et al.~\cite{bauer2021} or 
Hennig et al.~\cite{hennig2022}.

\noindent
In the first step, i.e., the model creation, the application developer defines a lattice-Boltzmann scheme by choosing the desired properties, such as the stencil, the collision space, the forcing scheme, and the relaxation rates. 
Subsequently, this high-level representation is automatically converted to an internal symbolical representation and mapped to a computational kernel. 
In this layer, the user can also define boundary conditions, macroscopic output, and communication schemes required for the \gls{mpi} parallelization. 
The compute kernels are subsequently passed to the \pystencils package \cite{bauer2019}. 
Here the symbolic representation is eventually transformed to actual code with back ends for C-code for CPUs and CUDA or OpenCL for GPUs.
A dedicated module in \walberla facilitates the integration of the produced kernels in \walberla applications.
Following this workflow in \lbmpy allows ro employ highly optimized compute kernels. 
One key advantage of this approach is that it naturally achieves
performance portability across different hardware architectures 
based on the exact symbolic representation and its systematic transformation to optimized code.

\subsection{\walberlawind} \label{sec:turbines}
\noindent
The present framework \walberlawind provides an \alm implementation for wind turbine applications with the \acrlong{lbm}. 
Built as a distinct framework on top of the multi-physics framework \walberla that provides the core \cfd routines, it is written in C++17 and CUDA in case of usage of GPUs. 
The codebase design focuses on modularity to enhance maintainability and productivity on CPUs and GPUs. 
Other critical aspects are performance and performance portability to enable large-scale simulations of wind turbines and complete wind farms with acceptable runtimes.
In this section, we will provide an overview of our new framework. 
We will address its software design, the resulting layers of abstraction, and the ensuing performance aspects.
\begin{figure}[ht]
  \centering
  \includegraphics[width=0.9\linewidth]{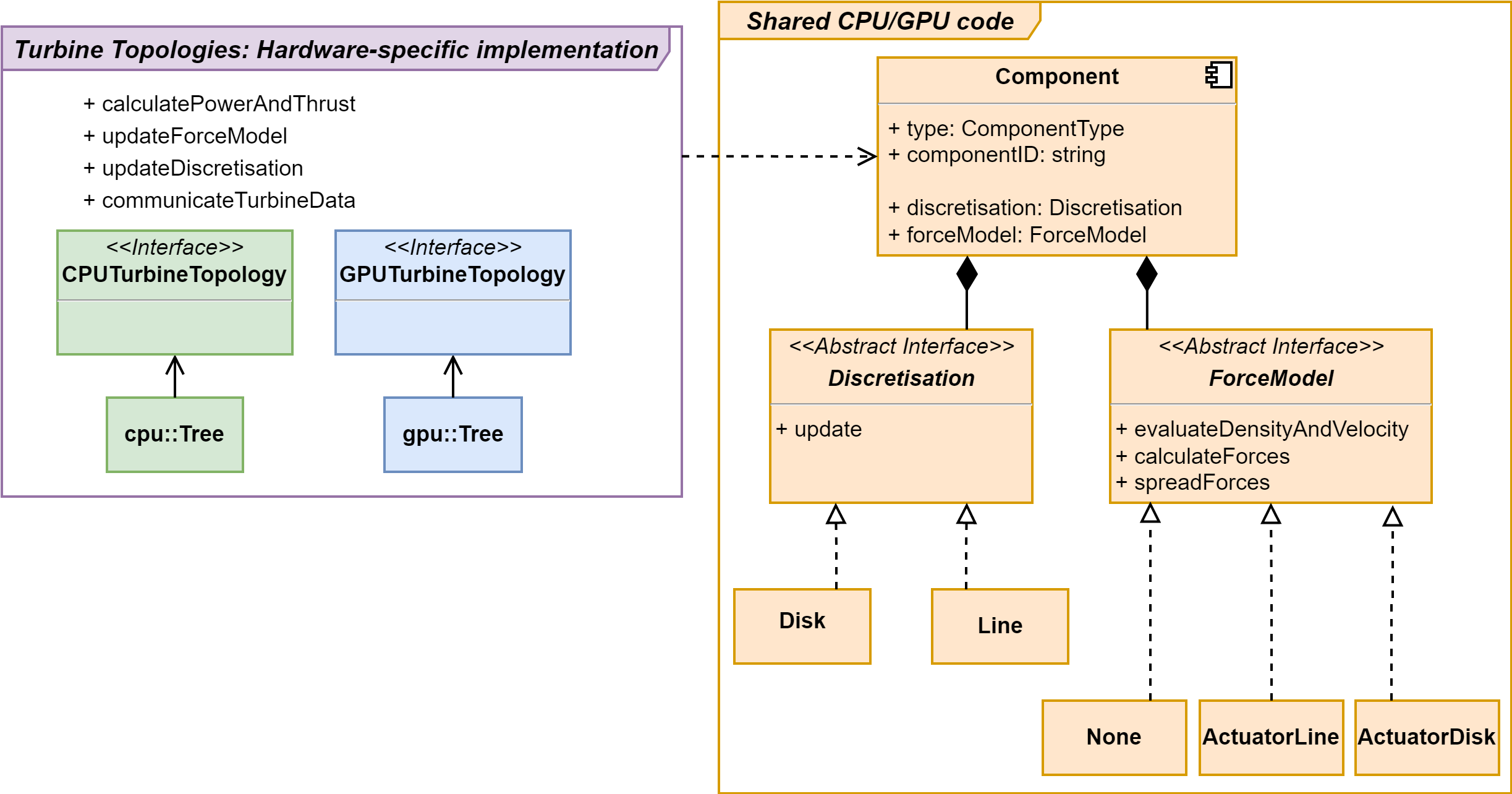}
\caption{UML diagram of the supporting structures and the physical modules. Pure CPU code (C++) is colored in green, pure GPU code (CUDA) in blue, and shared CPU/GPU code in yellow.}
\label{fig:class_uml}
\end{figure}
\paragraph{The turbine modules}
Three different module types build the framework: supporting structures, physical modules, and helper modules. 
\Cref{fig:class_uml} provides a simplified overview of the supporting structures, or \textsl{turbine topologies} and the physical models, or \textsl{components}.
The supporting structures define the general setup of a single wind turbine and how the distinct turbine parts are connected. 
The currently employed tree data structure that establishes a parent-child hierarchy between the components allows for flexible and versatile turbine configurations, e.g., classical \glspl{hawt}, \glspl{vawt}, and even floating wind turbines. 
This flexibility will be detailed later in Algorithms \ref{alg:update_discretisation_tree} and \ref{alg:update_discretisation_component}.
\Cref{fig:component_vs_topology} shows how different configurations lead to the same topology and therefore do not require special treatment on the user side.
\newsavebox{\imagebox}
\savebox{\imagebox}{\includegraphics[scale=0.225]{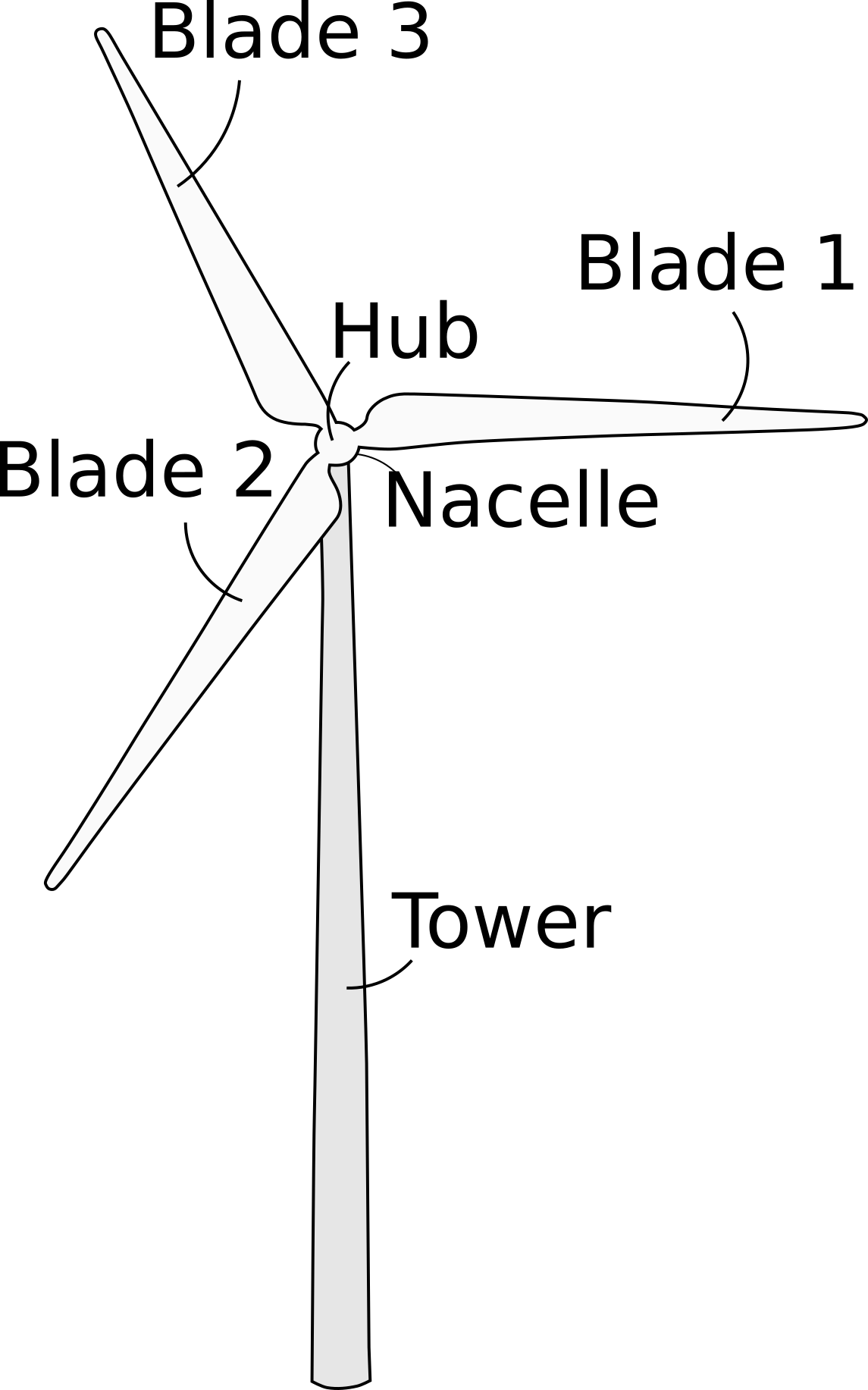}}
\begin{figure}[ht]
\begin{subfigure}[t]{.27\textwidth}
  \centering
  \usebox{\imagebox}
  \caption{Components of a \acrshort{hawt}}
  \label{fig:component_vs_topology:hawt}
\end{subfigure}
\hfill
\begin{subfigure}[t]{.26\textwidth}
  \centering
  \raisebox{\dimexpr.5\ht\imagebox-.5\height}{
  \includegraphics[scale=0.225]{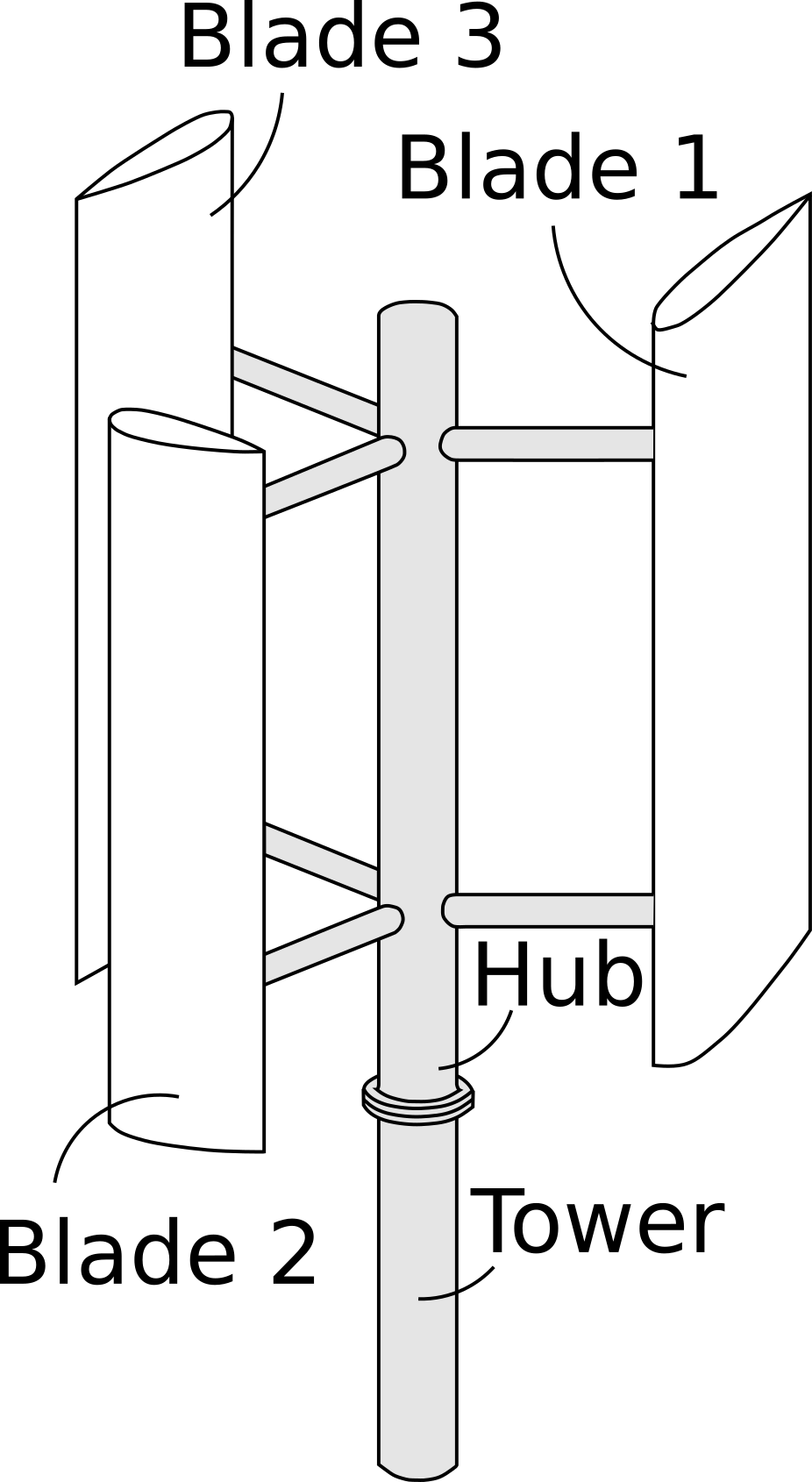} 
  }
  \caption{Components of a \acrshort{vawt}}
  \label{fig:component_vs_topology:vawt}
\end{subfigure}
\hfill
\begin{subfigure}[t]{.4\textwidth}
  \centering
  \raisebox{\dimexpr.5\ht\imagebox-.5\height}{
  \includegraphics[scale=0.225]{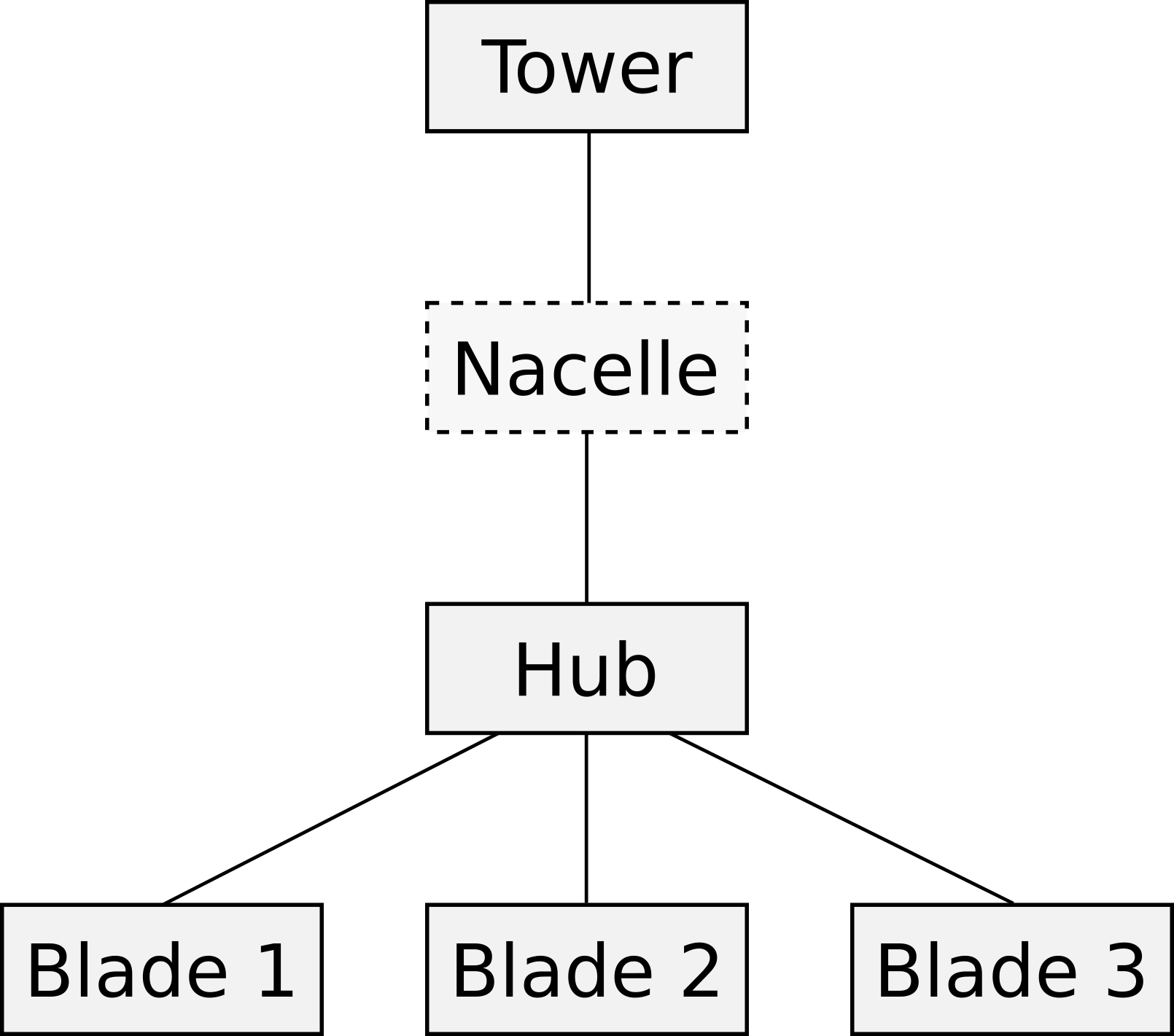}  
  }
  \caption{Topology of the turbines}
  \label{fig:component_vs_topology:tree}
\end{subfigure}
\caption{Comparison of the components of a \gls{hawt} and a \gls{vawt} in \subref{fig:component_vs_topology:hawt} and \subref{fig:component_vs_topology:vawt} and the topology view in our framework in \subref{fig:component_vs_topology:tree}. Despite the huge differences in geometry of the two wind turbines, their topology in our framework is identical apart from the absence of the nacelle in the \gls{vawt}.}
\label{fig:component_vs_topology}
\end{figure}
\noindent
Moreover, the supporting structures take care of all data I/O and communication needed in a simulation.
Especially when employing accelerating devices as GPUs, these routines can become very complex, as not only shared-memory communication on the host side with \gls{mpi} is required, but also host-to-device and device-to-host routines. 
Effectively, locating these crucial but complicated functions in the supporting structures, which a typical user uses as a black box, hides them and, therefore, enhances the usability and readability of physical modules.\\
In \Cref{fig:class_uml}, the distinction between hardware-specific code and shared code between CPUs and GPUs is clearly visible. 
In the example of hardware-accelerated simulations, the interface class \texttt{GPUTurbineTopology} is responsible for all the data handling in terms of memory management. 
In this setup phase of the simulation, it ensures that the read-in turbine data is transferred to the GPUs. 
During the simulation, it takes care of the inter-GPU communication and copies back data to the CPU in case of output desired by the user. 
Even the specific topology implementation, e.g.,  \texttt{gpu::Tree}, is free of memory handling routines and is mostly responsible for traversing the connectivity structure. Therefore, topology specializations are just as easily extensible for the end user as physical modules.
\\
The physical modules consist only of the actual turbine components, e.g., tower, nacelle, or blades. 
Every of which consists, in turn, of a particular discretization and force model. 
The former represents the turbine's spatial discretization and handles its update functions for turbine-global or component-local motion. 
All discretizations are defined in a component-local coordinate system and by the component's relative distance and angle to its parent in the employed hierarchy. 
This setup automatically propagates changes in the turbine geometry in a bottom-up fashion, which allows for flexibility in turbine configurations, as mentioned earlier. 
Algorithm \ref{alg:update_discretisation_tree} shows an example of the update routine of the geometry, i.e., the rotation of the blades for otherwise steady wind turbines, in a tree-like turbine topology. 
The update rules for specialized discretizations have to be implemented separately, e.g., Algorithm \ref{alg:update_discretisation_component}.
\begin{algorithm}
\caption{Algorithm of the discretization update for a tree-like turbine topology.}
\label{alg:update_discretisation_tree}
\begin{algorithmic}
\Function{Tree::updateDiscretization}{ }

discretization\_ $\rightarrow$ update() \Comment{call update function for this node's discretization; in our setup, it could be either a line or a disk}

\For{child \text{in} children\_} \Comment{call update for all children of this node}
    \State child $\rightarrow$ updateDiscretization()
\EndFor
\EndFunction
\end{algorithmic}
\end{algorithm}

\begin{algorithm}
\caption{Algorithm of the discretization update for disk and line discretizations.}
\label{alg:update_discretisation_component}

\begin{algorithmic}
\Function{Disk::update}{parental fix point}\Comment{Update routines for positions $\mathbf{p}$ and orientations $\mathbf{T}$ of specialized discretizations. $\{\cdot\}_p$ and $\{\cdot\}_c$ denote quantities of the parental fix point and component, respectively, $\{\cdot\}_r$ denote quantities relative to the parent. $\mathbf{R}$ is the rotation matrix obtained from the rotational velocity.}

    \State $\mathbf{T}_c \leftarrow \mathbf{T}_p \cdot \mathbf{T}_r \cdot \mathbf{R}$ \Comment{Update center point of disk.}
    \State $\mathbf{p}_c \leftarrow \mathbf{p}_p + \mathbf{T}_p \cdot \mathbf{p}_r$
\EndFunction

\vspace*{1em}

\Function{Line::update}{parental fix point}
    $\mathbf{T}^{start}_c \leftarrow \mathbf{T}_p \cdot \mathbf{T}_r \cdot \mathbf{R}$ \Comment{Update first point of the line.}
    $\mathbf{p}^{start}_c \leftarrow \mathbf{p}_p + \mathbf{T}_p \cdot \mathbf{p}^{start}_r$ 
    
    \For{$i$ \text{in} \#actuator points} \Comment{Update remaining points based on the start point.}
        \State $\mathbf{T}^i_c \leftarrow \mathbf{T}^{start}_c \cdot \mathbf{T}_r \cdot \mathbf{R}$
        \State $\mathbf{p}^i_c \leftarrow \mathbf{p}^{start}_c + \mathbf{T}^{start}_c \cdot \mathbf{p}^i_r$

    \EndFor

\EndFunction

\end{algorithmic}
\end{algorithm}
The second element of the component, the force models, implement the actuator routines for force calculation and spreading. 
Currently, we provide an implementation for actuator lines, with or without tip loss model and disks.
To clarify the difference between topologies and components, \Cref{fig:component_vs_topology} shows both on the example of a \gls{hawt} and a \gls{vawt}. 
While the geometry and the mode of operation differ heavily in actual wind turbines, they mostly consist of the same components, i.e., a tower, a hub, and blades, cf. \Cref{fig:component_vs_topology:hawt,fig:component_vs_topology:vawt}. 
Neglecting the \gls{vawt}'s stats, the \gls{hawt} only has an additional nacelle which the \gls{vawt} does not have. 
Displaying both turbines with their topology, \Cref{fig:component_vs_topology:tree}, shows that both configurations can be treated equivalently in our software and only differ by definition of the components' relative angles and distances.\\
Special support modules take care of remaining work. 
This consists of mathematical routines like interpolation schemes, the entire output handling, and all necessary data structures for the \walberla coupling.\\
A particular focus in \walberlawind lies in the extensibility and usability for the end-user. 
Therefore, all main modules, such as the turbine topologies, the discretizations, and the force models, were defined using an (abstract) interface against which all other modules link, see \Cref{fig:class_uml}. 
Thus, we ensure that the end-user can easily switch between specialized implementations without changing the remaining code. 
Adding new specializations is equally simple as no core structures 
must be adapted. 
Likewise, this enables the sharing of mutual code between them. 
While choosing the turbine topology has to happen at compile time in the main application, the choice of, e.g., force models can even happen at runtime depending on the final input file.\\
The level of abstraction discussed above has not only a crucial impact on usability and extensibility but, more importantly, also on performance portability. 
\lbm simulations are well-known to benefit highly from accelerating devices. 
Focusing on NVIDIA GPUs as accelerating devices in this work, we require CUDA as an additional programming language for GPU simulations. 
To fully exploit this potential in performance, it is only natural to have a GPU implementation of the \alm to minimize the host-to-device/device-to-host communication. 
The naive approach would be entirely separate implementations for CPUs and GPUs specific to the underlying programming language and hardware architecture. 
However, the principles and procedures in the physical modules do not change between CPU and GPU programming. 
Hence, our framework enforces a hardware-independent shared code base for all physical modules, as shown in the right-hand side of \Cref{fig:class_uml}.
For doing so, all routines in the physical modules can be processed by both CPU compilers and NVIDIA's \nvcc. 
Effectively, this means marking all functions as `\texttt{\_\_\_host\_\_ \_\_device\_\_}` when using the \nvcc and only using native data types, i.e., in particular, no STL-vectors or smart pointers.
As topology classes take care of all the hardware-specific data handling, this is sufficient to ensure the hardware-independent implementation of physical routines.
Additionally, we provide the same \gls{api} for CPU and GPU implementations. 
With this, we can can switch between different hardware simply, e.g., by using the class `\texttt{topology::\textbf{gpu}::Tree}` instead of `\texttt{topology::\textbf{cpu}::Tree}`. Furthermore, developers can extend the software back end with other programming models, e.g., HIP for AMD graphic cards, or OpenCL, by adding an additional topology class without the need to rewrite the code completely. 
Note, however, that this programming model must also be supported by \walberla.
\paragraph{Domain decomposition and communication}
As the turbine framework is only coupled to the fluid solver by a forcing term and not by the turbines' geometry, the domain decomposition remains almost as simple as in pure \lbm simulations. In contrast to fully resolved setups, our domain does not contain solid cells that create a workload imbalance but only fluid cells. However, processes that handle blocks containing turbines will undergo an increased computational load due to the cost of the actuator-type models and associated data interpolation and force projections.
To tackle the arising load-balancing issue, we rely on \walberla's pre-existing algorithms \cite{bauer2020}, in our case, a weighted Hilbert space-filling curve. Turbine blocks receive a higher weight to ensure a distribution among the processes as equal as possible. \\
While this domain partitioning ensures performance and scalability, it poses a challenge for the turbine module. 
With a partitioned domain, a wind turbine may spread forces over several blocks that even might reside on different processes. Therefore, they must exchange their data under certain circumstances with neighboring processes to ensure a consistent force spreading in the entire domain.\\
To minimize the amount of data during communication, we implement a strategy inspired by \cite{eibl2018}. 
This method applies to both CPU and GPU communication.
In the first step, the domain assignment, we construct a list of sub-domains that belong to a process and all its neighbors.
As the actuator points move in a predictable way, we can check in each time step if a local actuator point would project forces into a neighboring sub-domain or vice versa. If this is the case, we mark this point for communication. 
After identifying all critical actuator points, we pack them in a buffer, do the buffered communication and unpack the points on the corresponding processes. 
The force projection follows as usual. 
However, it is not limited to local actuator points as before but additionally treats those belonging to a neighboring subdomain that are spreading into the local domain.
With this approach, we minimize the amount of data while also minimizing the \gls{mpi} latency using buffered communication.\\
Lastly, we support two ways of inter-GPU communication. 
In many older computing systems, GPUs do not have the capability of directly exchanging data between them but have to use the system memory with CPU-side buffer memory to copy remote data. 
On newer systems, however, we optionally use NVIDIA's \textsl{GPUdirect} technology to circumvent the costly detour over CPUs but directly communicate all data between the GPUs.
Note that this technology is only employed for the communication of the \glspl{pdf} but not for the turbine data, as the latter's message sizes are too small to allow for efficient usage.


\section{Evaluation} \label{sec:results}
\noindent
The validation of our implementation follows the so-called \textsc{NewMexico} wind turbine test case.
This model scale wind turbine has been tested extensively in the large closed-walls low-speed DNW 9.5$\times$9.5$m^2$ wind tunnel test section in the context of the IEA Task 29, see \cite{mexnext2018}. 
In the present work, we focus on the axial flow conditions with three different \glspl{tsr}, the ratio between the blade tip velocity and the incoming wind velocity, as in \cite{mexnext2018}. 
These three \glspl{tsr}, approximately $10$, $6.6$ and $4$, are representative of the operating conditions of traditional wind turbines and translate to inflow wind velocities of $10$, $15$, and $24~m/s$. 
The wind turbine has a diameter of $4.5m$, with twisted and tapered blades based on three different airfoil profiles. More details, including operating conditions, can be found in \cite{mexnext2018}. 
Finally, we compare the blade force distribution and near-wake velocity profiles with experimental data and numerical results based on an inviscid free-wake lifting-line vortex flow solver called \castor \cite{blondel2017}. 
These solvers are known to predict the blade forces accurately and intrinsically account for tip losses.
%

\subsection{Case Setup} \label{sec:case_setup}
\subsubsection{\walberla setup}
\noindent
We simulate a fully periodic domain large enough to avoid wake re-entry. 
Moreover, the wind tunnel is large enough to neglect the effect of the surrounding walls, at least when considering blade forces and near-wake velocities, see \cite{rethore2011}. 
The domain is $25$ wind turbine diameters long and $5$ wind turbine diameters in height and width, which is assumed to be sufficient to avoid interactions when considering periodic domains. 
$50$ linearly distributed elements discretize the wind turbine blades. 
For a fair comparison with the vortex flow solver, we neglect tower and nacelle effects. 
In a pre-study, we considered three Mach numbers, i.e., $0.025$, $0.05$, and $0.10$. 
Unlike mentioned in \cite{asmuth2019}, no noticeable differences have been observed between Mach number of $0.05$ and $0.10$, while discrepancies emerge between Mach number of $0.10$ and $0.15$. 
Still, to ensure that presented results are converged in time, we ran all of the following simulations with the reference Mach number of $0.05$.
We employ different uniform grids with sizes ranging from $32$ up to $128$ lattice cells per wind turbine diameter to investigate the influence of the resolution. 
Note the better-resolved tip vortices and, consequently, blade tip losses using a mesh of $128$ lattice cells per wind turbine diameter. 
However, we also present the results of $64$ lattice cells per diameter to show the impact of the mesh resolution. 
The Glauert type tip-loss correction is deactivated to focus on the coupling between actuator-line models and \gls{lbm} and not on the specific AL implementations.
Furthermore, there is not yet a consensus in the wind energy community regarding tip-loss corrections for actuator-line simulations. 
Indeed, these corrections improve the results at high \gls{tsr} for coarse mesh but make results inconsistent with experiments at low \gls{tsr}, with fine meshes, or both.

\subsubsection{\castor setup}
\noindent
For the vortex solver \castor, the setup is slightly different due to the Lagrangian framework and is subject to previous convergence studies \cite{blondel2017}. 
The domain is unbounded, with no walls nor ground effect considered. 
The time step is larger and corresponds to a wind turbine rotation of $10$ degrees per time step. Each blade is discretized using $35$ cosine spaced so-called bound vortex filaments. 
The total simulation time is $15$ wind turbine rotations. 
Whereas actuator-line models cannot accurately represent the tip-loss due to the employed mesh sizes, vortex solvers do not need additional tip-loss corrections.


\subsection{Physical Validation} \label{sec:physical_validation}

\subsubsection{Blade forces distribution}
\begin{figure}[htbp]
\centering
\includegraphics[width=\textwidth]{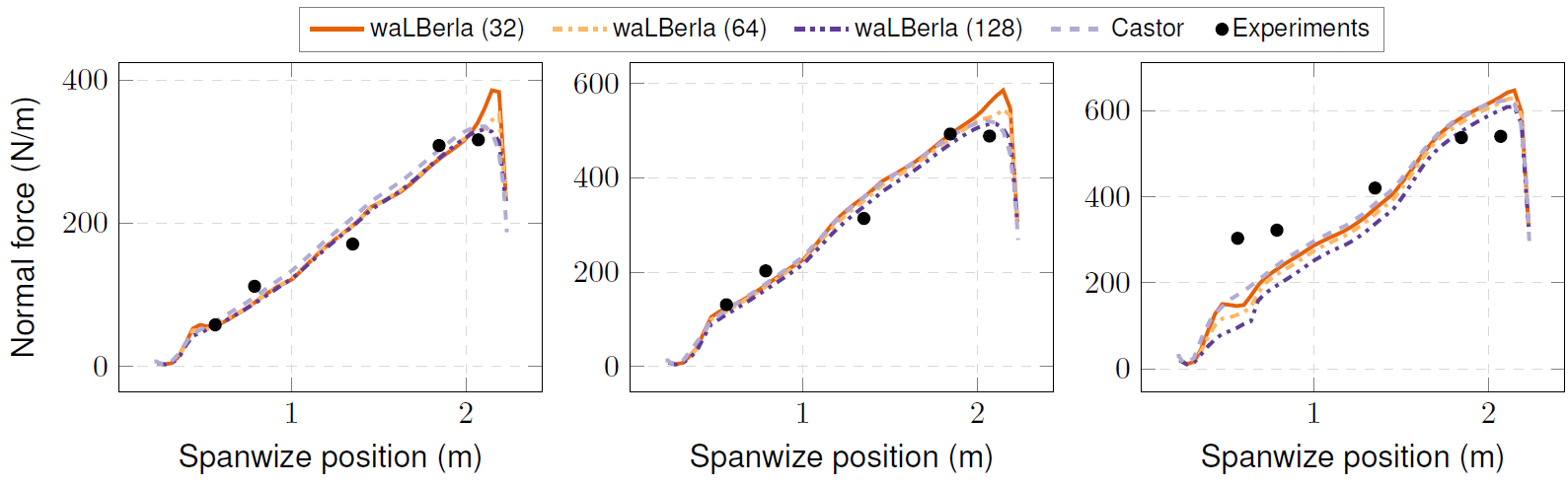}
\caption{Normal force distribution along the blade at approximately $10$ (left), $15$ (center) and $24~m/s$ (right).}
\label{fig:normal_forces}
\end{figure}

\begin{figure}[htbp]
\centering
\includegraphics[width=\textwidth]{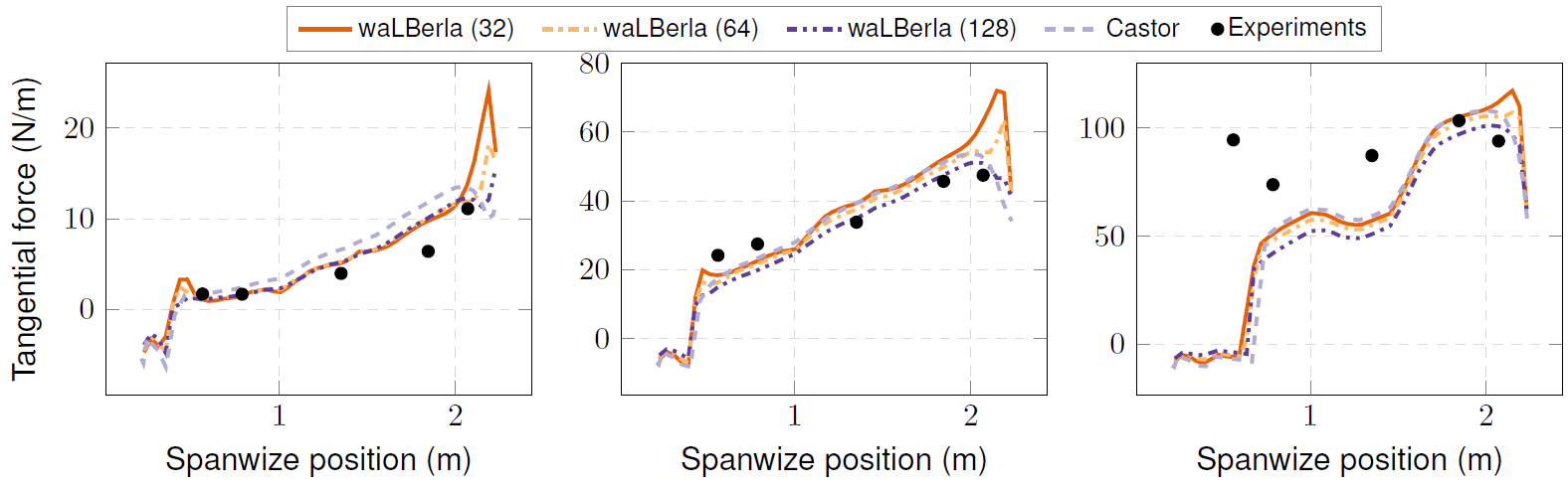}
\caption{Tangential force distribution along the blade at approximately $10$ (left), $15$ (center) and $24~m/s$ (right)}
\label{fig:tangential_forces}
\end{figure}

\noindent
\Cref{fig:normal_forces,fig:tangential_forces} show the force distribution along the blades for $32$, $64$, and $128$ lattice cells per wind turbine diameter. 
Normal forces are the forces experienced by the blade in the direction perpendicular to the local chord, while tangential forces act along the local chord. 
The solvers show a globally good agreement independent of the considered \gls{tsr}, an encouraging observation that validates our \gls{alm} implementation in \walberla. 

\definecolor{walberla-32}{RGB}{230,97,1}
\definecolor{walberla-64}{RGB}{253,184,99}
\definecolor{walberla-128}{RGB}{94,60,153}
\definecolor{castor}{RGB}{178,171,210}
\definecolor{experiments}{RGB}{0,0,0}

\pgfplotsset{
    stdaxis/.style={grid=major, grid style={dashed,gray!30}, legend style={nodes={scale=0.80, transform shape}}},
    walberla 32/.style={mark=none, color=walberla-32, style=very thick},
    walberla 64/.style={mark=none, color=walberla-64, style=very thick, dash dot},
    walberla 128/.style={mark=none, color=walberla-128, style=very thick, dashdotdotted},
    castor/.style={mark=none, color=castor, dashed, style=very thick},
    experiments/.style={only marks, color=experiments, style=very thick, mark size=1.5pt}
}

\newcommand{\legendspace}{\hspace*{0.3cm}}

\noindent
However, some discrepancies are noticeable near the tip of the blade. For the two lower wind velocities, i.e., higher \gls{tsr}, the vortex solver predicts a smoother distribution near the blade tip than the actuator-line models. 
In contrast to the actuator-line models, the vortex solver can accurately capture the blade tip vortices and the resulting tip losses. 
As shown in the figures, using a higher resolution settles this issue. 
While there is still a force peak at the blade tip for low resolutions (\walberla (32)), this peak decreases with higher resolutions (\walberla (64)) until the curve becomes smooth (\walberla (128)), and the force distribution tends towards the predictions from the vortex solver. 
However, the normal and tangential forces are globally decreasing when refining the mesh, especially at higher wind velocities. 
This phenomenon is not observed when using wider force-spreading kernels and needs further investigation in the future. 
Overall, the agreement with the experimental data can be considered satisfactory.  

\noindent
It is worth mentioning that within the \textsc{MexNext3} project, \cite{mexnext2018}, the validity of the measurements at mid-span was severely in dispute. 
Thus, comparisons with this specific experimental point must be taken with care.
Otherwise, the agreement between numerical results and experimental data is acceptable at the two lower wind velocities for normal and tangential forces. 
The tangential forces appear to be slightly overestimated in the second half of the blade, but the discrepancy remains small. 
The test case with the highest wind velocity is of particular physical interest. 
Indeed, the agreement between experimental data and numerical results is less good for the solvers. 
At such a high wind velocity, the angles of attack along the blade are high, above the static stall, leading to a low-pressure region on the blade's extrados. 
This low-pressure region and the blade rotation allow for the development of a spanwise flow. 
Actuator-line- or lifting-line-based-solvers cannot account for this highly three-dimensional phenomenon, leading to a significant disagreement between experiments and numerical results in the absence of analytical corrections.

\subsubsection{Near-wake velocities}
\begin{figure}[htbp]
\centering
\includegraphics[width=\textwidth]{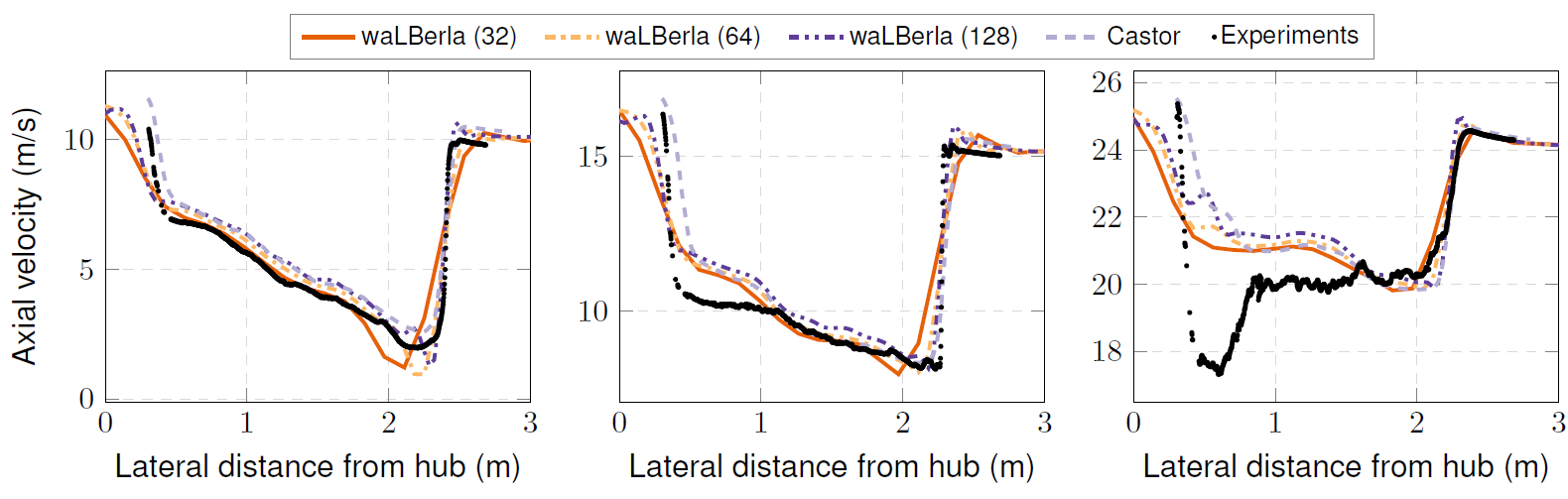}
\caption{Axial velocity profiles $30cm$ downstream of the wind turbine $10$ (left), $15$ (center) and $24m/s$ (right)}
\label{fig:rad_down}
\end{figure}

\noindent
Focusing on the very near-wake downwind velocity profiles, experimental measurements and predicted velocities match remarkably, see \Cref{fig:rad_down}. 
However, discrepancies emerge in the region close to the hub. 
This behaviour has already been reported extensively for actuator- and lifting-line-based approaches. 
These methods do not account for the three-dimensional flow patterns on the blade and also neglect hub and nacelle effects \cite{mexnext2018}.  
Despite this lack of an accurate physical model in this aspect when using actuator-line approaches, 
all numerical solvers are in good agreement with each other. 
One can notice that, as expected, velocity gradients near the tip and root of the blade get sharper when increasing the mesh resolution. 
Better resolved tip vortices lead to a better prediction of the tip losses.

\subsection{Performance} \label{sec:performance}
\noindent
Following the physical validation, another important aspect is to study and analyze computational performance.
Firstly, the absolute performance of a software program on a single CPU or GPU shows how well the code can exploit the particular compute resources. We will compare the measured \walberlawind performance on a single CPU/ GPU against the estimated performance of the pure \lbm kernel to see which overhead the turbines create.
Within the \gls{hpc} context, however, the scaling behavior of an application is of particular interest, as it indicates readiness for large-scale simulations. 
To this end, we investigate two different scaling setups, namely the \textit{strong} and the \textit{weak scaling}, for CPU, GPUs with a traditional inter-GPU communication scheme (\textit{GPU - regular MPI}), and GPUs with the GPUdirect technology (\textit{GPU - CUDA aware MPI}). 
The performance is reported in \gls{mlups}, a commonly used performance metric in lattice-Boltzmann simulations\cite{wellein2006}.
In both cases, we retain the \textsc{NewMexico} turbine specifics and fix the wind velocity to $15 \frac{m}{s}$, the Mach number to $0.1$ and the turbine resolution to $64$ lattice cells per diameter.
All simulations are run in \textit{single precision} on the Topaze supercomputer at CCRT/CEA.
Topaze has 864 compute nodes based on two AMD Milan@2.45GHz (AVX2) CPUs with 64 cores per CPU. 
Furthermore, it includes an accelerated partition with 48 compute nodes with four NVIDIA A100-SXM-80GB GPUs each.
\subsubsection{Single CPU/GPU performance}
A vital performance property of a software program is its absolute performance on a single compute instance, in our case, a single processor (CPU) or accelerator (GPU).
Especially in coupled multiphysics applications, however, this absolute performance metric can be hard to estimate and is not easily accessible. To this end, we will compare our measured performance for the wind turbine framework to the estimated peak performances of the pure \lbm code. As shown in \cite{bauer2019,holzer2020}, the optimized \lbm code generated by \lbmpy achieves almost peak performance on different hardware architectures. This comparison provides us, therefore, a suitable estimate of the overhead created by adding the wind turbine capabilities.
To assess the theoretical peak performance of the \lbm code, we follow the procedure proposed in \cite{holzer2020}.
The first step is to determine the limiting factor of the kernel, i.e., to establish whether the \lbm kernel is memory- or compute-bound. We introduce the machine balance $B_m$, the ratio of the memory bandwidth $b_{S}$ in \textsl{Bytes per second} ($B/s$) and the peak performance $P_{peak}$ in \gls{flops}, as well as the code balance $B_c$, the ratio of the data traffic $n_b$ and the number of floating-point operations $n_f$ in the kernel \cite{hager2017}. With these definitions, we can further compute the \textsl{lightspeed} of the kernel
\begin{align}
l = \min \left(1, \frac{B_m}{B_c} \right).
\end{align}
Memory-boundedness is defined by $l<1$, whereas the kernel is compute-bound for $l=1$. \\
For the NVIDIA A100 GPU, we obtained the peak single-precision (FP32) performance $P_{peak}=19.5$ T\gls{flops} from the vendor's datasheet. AMD, however, did not provide the AMD EPYC 7763's peak performance in its datasheet. Therefore, we will prove that the GPU kernel is memory-bound and assume the same for the CPU kernel. This assumption holds for most of \lbm kernels and should not influence the overall results. \\
The memory bandwidth $b_S$ can also be taken from the vendor's datasheet. Hager et al. \cite{hager2017} advise taking the bandwidth obtained by the STREAM benchmark \cite{McCalpin1995} as a reference instead, as several factors keep from reaching the maximum bandwidth provided by the vendor. The STREAM copy bandwidth measured on one A100 GPU of the Topaze supercomputer is $1713~GB/s$.\\
To obtain the data traffic, we need to consider the memory read and stored in one cell per iteration. In the following, we will assume the simplified best-case scenario where every cell value is loaded once and, otherwise, reused from the cache. For the applied $D3Q27$, we have to read and store 27 \gls{pdf} values per cell and timestep. Furthermore, we need to load a 3-dimensional force vector, i.e., three values in addition. Overall, we have $n_b = (27 + 27 + 3) * 4B = 228~B$ of data traffic for single-precision. \\
Lastly, we determine the number of floating-point operations per cell and iteration $n_f$. Instead of manually counting the number of operations by hand, we rely on \textsc{lbmpy}'s \texttt{count\_ops} functionality. With this, we obtain for the $D3Q27$ cumulant \lbm kernel $n_f = 828$ \acrshort{flop}.\\
Finally, the lightspeed of the pure \lbm kernel on the A100 GPU reads 
\begin{align}
l &= \min \left(1, \frac{B_m}{B_c} \right) = \min \left( 1, \frac{b_S \cdot n_f}{P_{peak} \cdot n_b} \right) = 0.3190 < 1.
\end{align} 
The lightspeed being significantly less than $1$, we can conclude that the GPU kernel is, indeed, memory-bound and that it is safe to assume likewise for the CPU kernel. With this information, the estimated maximum performance of the kernels in GLUPS is
\begin{align}
P_{max} = \frac{b_S}{n_b}.
\end{align}
We compare the measured performance of the wind turbine application, with and without turbines present, on a single CPU/ GPU against this estimated maximum performance. Table~\ref{tab:abs_performance} provides an overview of the measured data and gives the ratio of the performances.
\begin{table}
    \centering
    \resizebox{\textwidth}{!}{
    \begin{tabular}{|c|c|c|c|c|c|c|}
        \hline
        Hardware & $n_b$ in $\frac{B}{FLOP}$ & $b_S$ in $\frac{GB}{s}$ & \makecell{Estimated perf.\\ $P_{max}$ in MLUPS} & \makecell{measured perf. without\\ turbines in MLUPS} & \makecell{measured perf. with\\ turbines in MLUPS} & $\frac{\text{Perf. with turbines}}{P_{max}}$ in $\%$ \\
        \hline
        AMD Epyc 7763 & 228 & 105.2 & 461.4 & 204.0 & 202.1 & 43.8 \\
        NVIDIA A100 & 228 & 1713.0 & 7513.2 & 1866.5 & 1677.0 & 22.3 \\
        \hline
    \end{tabular}
}
    \caption{Estimated and measured kernel performance for a resolution of $64$ cells per diameter.}
    \label{tab:abs_performance}
\end{table}
Holzer et al. \cite{holzer2020} reach at least $82\%$ of the theoretical peak performance on NVIDIA P100 and V100 cards for pure \lbm kernels generated in double-precision by \lbmpy. In our application with wind turbines, we only reach $43\%$ on a single CPU with 64 cores and $22\%$ on a single GPU. 
To investigate the origin of this substantial loss in performance, we also ran the exact same simulation setup with zero wind turbines. Table \ref{tab:abs_performance} shows that this measure improves the performance marginally, i.e, from 202 to 204 \gls{mlups} on the CPU and from 1677 to 1866 \gls{mlups} on the GPU. We can conclude that despite its complexity, the wind turbine module does not introduce substantial overhead. 
In future work, it remains to explore possible optimizations to achieve the performance ratios reported by Holzer et al. Also, the influence of the differing floating-point precisions needs to be studied. \\
For the simulations on GPUs, one factor is surely the occupancy of the GPU. In the following strong scaling results, Figure \ref{fig:strong_scaling}, we see that increasing the resolution from $64$ to $128$ cells per diameter, i.e., a factor of $8$ in workload, almost doubles the performance on one node. A single GPU does not support the memory requirement for this test case with $128$ cells per diameter. However, a similar behavior is expected as for one node, potentially increasing the performance ratio to $44\%$.
\subsubsection{Strong scaling}
\noindent
Strong scaling studies the speed-up for a fixed problem size with respect to the number of computational units. 
Here, we fixed the domain size corresponding to the one used for the physical validation, i.e., $25 \times 5 \times 5$ turbine diameters, and only modified the number of processors. The load balancing issue is circumvented by assigning one block per process.\\
\Cref{fig:strong_scaling} (left) shows the scaling results for a resolution of $64$ lattice cells per diameter. 
The CPU runs exhibit excellent, nearly linear scaling behavior that flattens only slightly for higher node counts. 
The GPU runs, however, do not show this favorable trend. On the positive side, we expectedly observe a beneficial effect of the GPUdirect technology on the performance and even a fairly linear curve until $5$ nodes or $20$ GPUs. 
Yet, once we exceed $5$ nodes, we obtain a performance plateau with no significant speed-up. 
Such behavior is clearly a result of the under-utilization of the GPUs. 
When the workload per GPU becomes too small, adding more GPUs will make the workload even smaller and will thus lead to a further increased overhead that prohibits scaling. 

 \pgfplotsset{
    stdaxis/.style={grid=major, grid style={dashed,gray!30}, legend style={nodes={scale=0.80, transform shape}}},
    CPU/.style={mark=triangle*, color=walberla-128, style=very thick, every mark/.append style={solid}},
    GPU normal/.style={mark=square*, color=walberla-32, style=very thick, every mark/.append style={solid}},
    GPU direct/.style={mark=*, color=walberla-64, dashed, style=very thick, every mark/.append style={solid}},
}

\begin{figure}[htbp]
\centering
\begin{tikzpicture}
\begin{groupplot}[
  group style={
    group size=2 by 1,
    vertical sep=0pt,
    group name=performance_strong,
	xlabels at=edge bottom,
	ylabels at=edge left
    },
  width=0.5\textwidth,height=0.3\textwidth,
  xlabel={Number of compute nodes}, 
  ylabel={MLUPS},
  y tick label style={/pgf/number format/sci subscript},
  stdaxis
  ]

\nextgroupplot[xmin=-1, xmax=31, ymin=0]
\addplot [GPU direct] table [x index=0, y index=1]{strong_scaling_NewMexico_64diam_MLUPS.dat};
\addplot [GPU normal] table [x index=0, y index=2]{strong_scaling_NewMexico_64diam_MLUPS.dat};       
\addplot [CPU] table [x index=0, y index=3]
{strong_scaling_NewMexico_64diam_MLUPS.dat};

\nextgroupplot[xmin=-1, xmax=31, ymin=0, legend to name={PerformanceLegend_strong}, legend style={legend columns=3,draw=gray}]
\addplot [GPU direct] table [x index=0, y index=1]{strong_scaling_NewMexico_128diam_MLUPS.dat};
\addplot [GPU normal] table [x index=0, y index=2]{strong_scaling_NewMexico_128diam_MLUPS.dat};       
\addplot [CPU] table [x index=0, y index=3]
{strong_scaling_NewMexico_128diam_MLUPS.dat};  

\addlegendentry{GPU - CUDA aware MPI\legendspace}
\addlegendentry{GPU - regular MPI\legendspace}
\addlegendentry{CPU\legendspace}

\end{groupplot}

\path (performance_strong c1r1.north east) -- node[above]{\ref{PerformanceLegend_strong}}+(0,0.75cm) (performance_strong c2r1.north west);
\end{tikzpicture}
\caption{Computational performance measured in \gls{mlups} for the strong scaling experiment for resolutions of $64$ cells per diameter (left) and $128$ cells per diameter (right). Each node consists of $128$ CPU cores or $4$ GPUs, the domain size is $25\times 5 \times 5$ turbine diameters.}
\label{fig:strong_scaling}
\end{figure}
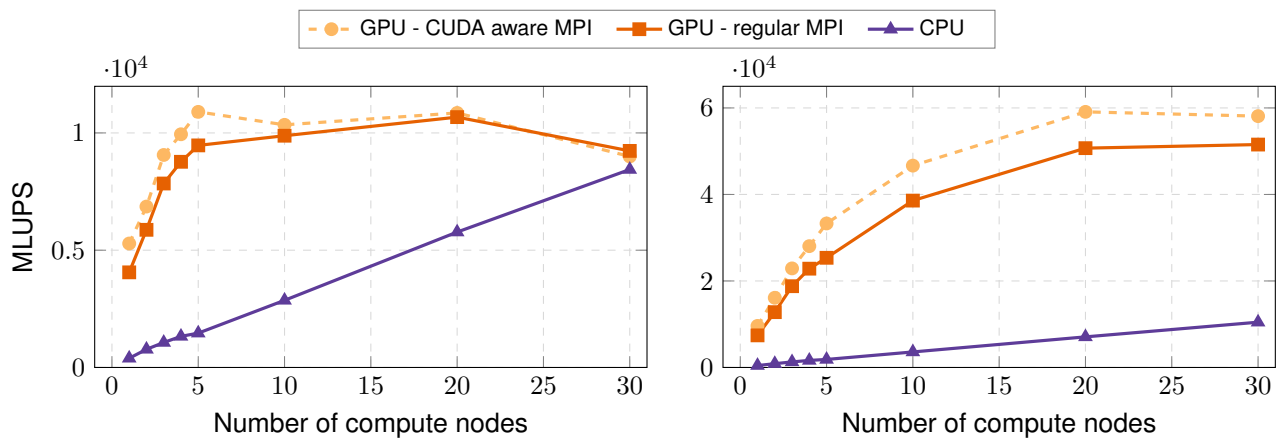
\noindent
To confirm this hypothesis, we repeated the scaling runs with a higher resolution of $128$ lattice cells per diameter, thus increasing the total problem size by $8$. The corresponding results are visualized in \Cref{fig:strong_scaling} (right). Again, the CPU results are excellent, showing a slight improvement in the overall performance and speed-up. 
Notably, the overall trend on GPUs is now significantly different. 
The fairly linear curve until $5$ nodes is retained, but now with significantly higher performance in \gls{mlups}, improving from less than $10$ to less than $30$ GLUPS. 
For more than $5$ nodes, the plateau is now transformed into a saturation curve typical for strong scaling experiments. 
Also worth mentioning is the persistent beneficial impact of the GPUdirect technology for the larger scaling run, i.e., a speed-up of up to 31.3\% using $5$ nodes.
In a simplified analysis with \textit{Amdahl's lae}, we can argue that the maximum speed-up is defined by the serial part of an application. Asymptotically, this limit is approached when adding more computational resources. 
It is unavoidable to reduce the fraction of serial execution to increase the maximal speed-up.
In particular, this includes all \gls{mpi}-related routines, i.e., the communication of the \glspl{pdf}, the communication of the forces, and the communication of the wind velocity between processes. While the application spends about $25\%$ of its runtime in such routines for $1$ node, this percentage increases to about $65\%$ for $30$ nodes. This increase is partially due to the speed-up in compute-intensive routines. But the absolute time in $s$ spent in communication also increases significantly.

\subsubsection{Weak scaling}
\noindent
In weak scaling, we  study the speed-up for a scaled problem size with respect to the number of computational units. 
In contrast to strong scaling, which is an indicator of how many computational resources one should assign to the simulation of a fixed domain, the weak scaling scenario may be more characteristic in the context  of large scale wind energy simulations. 
While it is important to solve problems of a certain size as efficiently as possible, in wind energy science, we are also interested in the simulation of growing problem sizes, e.g., for the simulation of wind farms.\\ 
For this setup, we assign a subdomain of $260\times 260\times 260$ lattice cells per processor, thus scaling with the number of computational resources. 
Here it is worth mentioning that for typical weak scaling experiments, not only the domain size but also the number of turbines needed to be scaled. Since our turbine implementation is not yet fully distributed amongst processors, we chose to keep the number of turbines in the domain constant, i.e., one turbine with a resolution of $64$ lattice cells per diameter. 
Here we focus on demonstrating the weak scaling behavior of the core solver, as physically validated above, while the parallelization and validation of models with multiple wind turbines will be subject of future work. 

\begin{figure}[htbp]
\centering
\begin{tikzpicture}
\begin{groupplot}[
  group style={
    group size=1 by 1,
    vertical sep=0pt,
    group name=performance_weak,
	xlabels at=edge bottom,
	ylabels at=edge left
    },
  width=0.5\textwidth,height=0.3\textwidth,
  xlabel={Number of compute nodes}, 
  ylabel={MLUPS per GPU},
  y tick label style={/pgf/number format/sci subscript},
  stdaxis
  ]

\nextgroupplot[xmin=-1, xmax=33, ymin=0,ymax=400,legend to name={PerformanceLegend_weak}, legend style={legend columns=3,draw=gray}]
\addplot [GPU direct] table [x index=0, y index=2]{weak_scaling_NewMexico_64diam_MLUPS.dat};  

\addlegendentry{GPU - CUDA aware MPI\legendspace}

\end{groupplot}

\end{tikzpicture}
\caption{Computational performance measured in \gls{mlups} for the weak scaling experiment. Each node consists of $4$ GPUs, and each processor obtains a subdomain of $260\times 260 \times 260$ lattice cells.}
\label{fig:weak_scaling}
\end{figure}
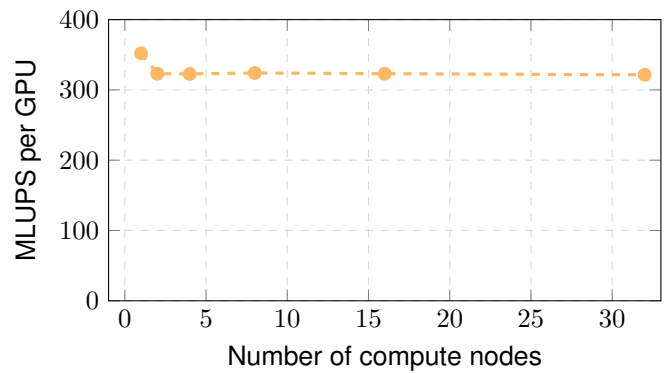
\noindent
\Cref{fig:weak_scaling} shows the number of \gls{mlups} per GPU, up to $30$ nodes or $120$ GPUs under weak scaling conditions.
Noticeably, there is a small jump in performance between one and two nodes, most likely due to the shift from inter-node communication to intra-node communication. Despite this jump, the performance per GPU stays quasi-constant as predicted by \textit{Gustafson's law}. 
With more than 17.5 million cells per GPU, we execute on average 74.46 time steps per second during the weak scaling experiment.
Achieving scalability to such high node counts indicates the capability of \walberlawind to deal with larger domains and eventually wind farms.


\section{Conclusion} \label{sec:conclusion}
\noindent
In the present study, we introduced our strategy, methods and software to combine the \acrfull{alm} and the \acrfull{lbm} for the simulation of wind turbines. While the \gls{alm} has been employed for decades in wind energy science, the \gls{lbm} only started to become interesting for those high-Reynolds number flows in the last years due to the development of new, more robust collision operators. Recent publications \cite{asmuth2020general,asmuth2019,asmuth2020compressibility} already performed extensive studies on the overall suitability of \gls{lbm} for wind turbine simulations. However, these studies were executed only on single-node machines. To our best knowledge, this article proposes the first implementation for large-scale systems with a particular focus on multi-node performance.\\
We presented the computational framework used for our studies, namely \walberla, \lbmpy, and our new code base \walberlawind. An important aspect of \walberlawind's software design is its modularity, with which we ensure excellent extensibility and performance portability between CPUs and GPUs. 
Splitting up the routines into \textit{physical modules} and supporting structures, the \textit{turbine topologies} allows for a separation of concerns: the physical modules share the same code between different hardware architectures and are only responsible for calculating and performing the respective sub-routines. The turbine technologies, however, are realized hardware-specific and take care of all data handling and communication. This approach avoids code divergence and facilitates the implementation of new features and physical models.\\
An analysis based on the \textsc{NewMexico} test case was performed for the blade force distribution and the near wake velocities and compared to experimental results and an inviscid free-wake lifting-line vortex solver. Different resolutions in our setup and different \acrfull{tsr}s were investigated. For higher \glspl{tsr}, we obtained excellent results, matching those of the experiments and the reference solver. For lower \glspl{tsr}, \walberlawind still matched well with the reference solver but showed stronger deviations from the experimental results. This observation is due to the presence of spanwise flows along the blade that the present numerical methods can inherently not account for without analytical corrections, as well as neglected hub and nacelle effects.\\
Performance measurements on a single CPU and a single GPU showed that there is still a drastic gap between the application and its theoretical peak performance that is mostly attributed to a non-optimal setup of the \lbm module. Possible optimization measures will be part of future work.
Lastly, we performed strong and weak scaling experiments to investigate the suitability of our code base for large-scale machines. We compared CPU and GPU runs and studied the influence of \textsc{Nvidia}'s GPUdirect technology on the performance. Our strong scaling runs with a resolution of $64$ lattice cells per diameter point out the importance of occupancy of GPUs. While the CPU runs scale almost linearly, the GPUs soon reach a performance plateau. Choosing a higher resolution, i.e., a bigger problem size, this problem is overcome, and the typical saturating behavior for strong scaling is obtained. In the weak scaling scenario, we could show that our code yields almost constant performance per GPU, indicating excellent suitability for large-scale simulations.
Note that the performance in both scaling cases can be further improved by reducing the serial fraction of our code. In particular, the turbine implementation is currently not fully distributed and consequently increases the serial fraction. The full parallelization of the turbine models will be addressed in future work.
Moreover, we will study the impact of load balancing when introducing more blocks than processes, as it is typical in mesh refinement applications.\\
In summary, we believe that our implementation exhibits great potential for future use in the large-scale, high-performance simulation of wind turbines and entire wind farms. 


\section*{Acknowledgements} \label{sec:acknowledgements}
\noindent
This project has received funding from the European Union’s Horizon 2020 research and innovation
programme under grant agreement No 824158 (EoCoEII), as well as the Deutsche Forschungsgemeinschaft (DFG, German Research Foundation) under project No 433735254.

\end{doublespacing}

\section*{References}

\begingroup
\renewcommand{\section}[2]{}
\bibliographystyle{unsrt}
\bibliography{literature}
\endgroup
    
\end{document}